\documentclass[12pt]{article}

\usepackage{amsmath}
\usepackage{amssymb}
\usepackage{amsfonts}

\def\out#1{}

\def\D{{\cal D}}
\def\Re{\textrm{Re}}
\def\vol{\textrm{vol}}
\def\vol{\mathit{Vol}}

\def\diag{\textrm{diag}}
\def\d{{\rm d}}
\def\im{{\rm i}}
\def\fd#1#2{\frac{\delta#1}{\delta#2}} % functional derivative

\def\K{{\cal K}}

\def\hhat{\hat{h}}
\def\HH{{\rm H}}
\def\H{{\cal H}}
\def\Hhat{\hat{{\cal H}}}
\def\HHhat{\hat{H}}
\def\HHHhat{\hat{\HH}}
\def\J{{\cal J}}

\def\R{{\cal R}}

\def\ave#1{\left\langle#1\right\rangle}
\def\av#1{\langle#1\rangle}

\def\ang#1#2{\langle#1,#2\rangle}

\def\SQ{S_\textrm{Q}}
\def\sch{Schr\"odinger}
\def\kuchar{Kucha\v{r}}

\begin{document}

\title{Nonlinear quantum gravity\\ on the constant mean curvature foliation}

\author{Charles H-T Wang}

\date{}

\maketitle

\begin{center}
\vskip -8mm
%\paint{
Department of Physics, Lancaster University, Lancaster LA1 4YB, UK\\
(until 31 December 2004)\\
~\\
School of Engineering and Physical Sciences,\\
University of Aberdeen, King's College, Aberdeen AB24 3UE, UK\\
(starting 1 January 2005)\\
%(new permanent address starting 1 January 2005)\\
%~\\
%Centre for Fundamental Physics, Rutherford Appleton Laboratory,\\
%Chilton, Didcot, OX11 0QX, UK\\
%~\\
%E-mail: charles@wang.me.uk
%}

% (gr-qc/0406079)
\end{center}

\abstract{A new approach to quantum gravity is presented based on
a nonlinear quantization scheme for canonical field theories with
an implicitly defined Hamiltonian. The constant mean curvature
foliation is employed to eliminate the momentum constraints in
canonical general relativity. It is, however, argued that the
Hamiltonian constraint may be advantageously retained in the
reduced classical system to be quantized. This permits the
Hamiltonian constraint equation to be consistently turned into an
expectation value equation on quantization that describes the
scale factor on each spatial hypersurface characterized by a
constant mean exterior curvature. This expectation value equation
augments the dynamical quantum evolution of the unconstrained
conformal three-geometry with a transverse traceless momentum
tensor density. The resulting quantum theory is inherently
nonlinear. Nonetheless, it is unitary and free from a nonlocal and
implicit description of the Hamiltonian operator. Finally, by
imposing additional homogeneity symmetries, a broad class of
Bianchi cosmological models are analyzed as nonlinear quantum
minisuperspaces in the context of the proposed theory.}


\vskip 10mm

\section{Introduction}

The term `quantization' is ill defined. Ideally, one starts with a
quantum theory and may wish to classicize it. However, until the
ultimate quantum theory of everything is found, one continues to
`prepare' a classical system for quantization under a consistent
scheme. Further justifications of the quantized system may then be
sought. In the canonical approach, this requires a choice of phase
variables. Gauge or other auxiliary degrees of freedom, if any,
must be factored out from the redundant phase space so that only
the physically significant degrees of freedom are quantized. The
corresponding type of variables are often referred to as {\it
dynamical variables} whose identification from a given theory is
not always straightforward and can be ambiguous
\cite{BergmannGoldberg1955}. Thus, a central question to ask in
quantizing a classical system is: {\em what is the right point of
departure for quantization?} Once this is decided, the actual
quantization may then be addressed in terms of its physical
consequence and mathematical expediency. The problem of
quantization is particularly perplexing for gravity. Starting with
the Dirac-Arnowitt-Deser-Misner (Dirac-ADM) formalism of general
relativity \cite{ArnowittDeserMisner1962}, the implementation of
canonical quantization has been impeded by the presence of the
Hamiltonian and momentum constraints that signal the redundancy of
the conventional geometrodynamical variables. While the momentum
constraints are clearly related to the invariance of the formalism
under spatial diffeomorphisms, a gauge interpretation of the
Hamiltonian constraint is more evasive. In particular, the Dirac
algebra satisfied by these gravitational constraints fails to be a
Lie algebra of any gauge group.

Two divergent strategies have been adopted in dealing with the
gravitational constraints. The first strategy aims to exploit them
as {\it first class constraints} to generate the quantum dynamics
of the spatial geometry through the {\em Dirac constraint
quantization}  \cite{Dirac1950, Dirac1958a, Dirac1964,
GotayNester1984, SeilerTucker1995}.  In essence, this scheme turns
both Hamiltonian and momentum constraints into quantum
annihilation equations for the physical state functional. The
second strategy is based on the {\em reduced phase space} method,
whose basic theory and history may be found in
\cite{MarsdenWeinstein1974, Marsden1981, MarsdenRatiuScheurle2000,
MarsdenWeinstein2001, SniatyckiWeinstein1983, IsenbergMarsden1984,
PonsShepley1995, PonsShepley1995a}. When applied to canonical
general relativity this strategy advocates the elimination of all
gravitational constraints in order to obtain a nonvanishing
effective Hamiltonian that generates the evolution of certain
unconstrained geometric variables
\cite{ChoquetBruhatFischerMarsden1979, IsenbergMarsden1982,
Isenberg1987, Gotay1986}. The reduced system
%is then
would be
amenable to the conventional \sch{} quantization in principle.

The Dirac constraint quantization strategy gains support from the
confirmation that the Dirac algebra guarantees the
hypersurface-independence of the dynamic evolution and hence the
general covariance of any parametrized field theories
\cite{Teitelboim1973, HojmanKucharTeitelboim1973}. Together with
the interconnection between the Hamiltonian and momentum
constraints \cite{Kuchar1981} this property suggests that, under
Dirac constraint quantization, a hypersurface-independent quantum
evolution of gravity could be constructed without the need for
gauge fixing. As the the Hamiltonian is quadratic in the canonical
momenta of the spatial metric, this strategy results in the
Wheeler-DeWitt equation \cite{Wheeler1962, DeWitt1967}, which is a
functional Klein-Gordon equation for three-geometries that has no
obvious probabilistic interpretation. While this non-unitary
quantum evolution may lead to interesting predictions such as the
tunnelling phenomena in certain models of the early universe
\cite{Vilenkin1984}, considerable efforts, notably by \kuchar{}
\cite{Kuchar1971, Kuchar1972, Kuchar1992}, have been made to find
alternative set of gravitational variables in which Dirac
quantization yields the Tomonaga-Schwinger type equations
\cite{Tomonaga1946, Schwinger1948} that generate unitary quantum
evolution of geometry. For a broad class of midisuperspace models
with two commuting spatial isometries, the symmetry-reduced
general relativity can be shown to be equivalent to a parametrized
scalar doublet field theory in 2-dimensional Minkowski space
\cite{Torre1999}. However, it is not clear whether \kuchar's
variables can be found in full general relativity
\cite{Kuchar1978}. Further discussion on \kuchar's method may be
found in \cite{WangKessariIrvine2004} and references therein.

The strategy of applying the \sch{} quantization to a reduced
gravitational phase space by means of gauge fixing was initially
explored by Dirac \cite{Dirac1959} and later by ADM
\cite{ArnowittDeserMisner1962}.  The ADM procedure amounts to
choosing four out of the six spatial metric components as special
spacetime coordinates. The gravitational constraints are solved
(and are thereby eliminated) for the canonical momenta of these
coordinates, leading to a nonvanishing `true' Hamiltonian. The
remaining two spatial metric components with the corresponding
momenta are then regarded as dynamical variables. Note that this
procedure does not by itself single out any preferred dynamical
variables. The analogy that ADM drew between the gravitational and
electromagnetic fields led them to introduce a background
Minkowski metric, with respect to which the transverse traceless
part of the general spatial metric components and their momenta
are treated as dynamical variables to be quantized. However this
scheme relies on an {\it ad hoc} reference metric at the expense
of loosing covariance.

In this regard, a much more satisfactory gauge fixing scheme that
singles out a `spin-2' part of gravitational field was developed
by York \cite{York1971, York1972, York1973}. Built upon previously
works by Lichnerowicz, Brill and Choquet-Bruhat
\cite{Lichnerowicz1944, Brill1959, ChoquetBruhat1962}, York
defined a decomposition of general second rank symmetric tensors
of any weight into transverse traceless, longitudinal and trace
parts, in a conformably covariant fashion. The application of this
powerful scheme to canonical general relativity is most revealing
if the time-slicing endows each spatial hypersurface with a
constant mean exterior curvature. Under this constant mean
curvature (CMC) foliation, the spatial metric is written in terms
of the scale factor (of the volume element) and a conformal
metric. The latter is used to decompose the metric momentum
tensor. The resulting longitudinal part of the metric momentum
tensor must vanish to satisfy the decoupled momentum constraints.
The trace part of this tensor is proportional to the mean exterior
curvature which labels the spatial hypersurfaces like time. This
`exterior time' may also be considered as a manifestation of the
geometric carrier of information about time originally suggested
by Baierlein, Sharp and Wheeler \cite{BaierleinSharpWheeler1962}.
Therefore, the transverse traceless part of the metric momentum
tensor, along with the conformal metric, naturally carries
dynamical degrees of freedom and can be interpreted as
gravitational waves \cite{York1973}. The scale factor is to be
determined by solving the reduced Hamiltonian constraint as an
elliptic equation, i.e. the Lichnerowicz equation
\cite{OMurchadhaYork1973, ChoquetBruhatYork1980}. Furthermore, it has been shown by
Isenberg, O'Murchadha, York and Nester that York's conformal
treatment is capable of accommodating scalar, electromagnetic,
spinor fields \cite{IsenbergOMurchadhaYork1976} and indeed
`virtually all' known physically relevant forms of matter field
\cite{IsenbergNester1977}.

The clear geometrical meaning and physical interpretation of
York's identification of the dynamical variables of general
relativity raised the hope for quantum gravity by quantizing the
conformal three-geometry with transverse traceless momentum. The
diffeomorphisms of the three-geometry and the longitudinal part of
the metric momentum (nullified by solving the momentum constraints
using the CMC condition) may be regarded as gauge variables. In
contrast, no simplicity seems to exist in dealing with the
remaining Hamiltonian constraint. If this constraint is also
solved for the scale factor before quantization as per the ADM
procedure, then a nonvanishing Hamiltonian for canonical general
relativity may be constructed, which turns out to be the volume
integral of the universe. As a nontrivial elliptic equation must
be solved, this Hamiltonian is a nonlocal functional of the
dynamical variables and extrinsic time and is only defined
implicitly. This somewhat disturbing feature
has been noted by Choquet-Bruhat, York,
Fischer and Moncrief \cite{ChoquetBruhatYork1980,
FischerMoncrief1997, FischerMoncrief1999, FischerMoncrief2000a}.
It presents at least a technical obstacle to quantization as
issues such as factor ordering ambiguities are hard to tackle with
an implicit and nonlocal Hamiltonian \cite{Carlip2001}.

Under this state of the affair, alternative quantization schemes
should not be ruled out where York's identification of the dynamic
part of gravity may be utilized without having to solve the
Hamiltonian constraint prior to quantization. The purpose of this
paper is to explore one such possibility. Specifically it is
proposed that the scale factor may be treated on an equal
footing with the lapse function as a Lagrangian multiplier as
befits their roles in formulating the kinematics of the spatial
hypersurfaces. These non-dynamical variables are conceptually akin
to Dirac's surface variables in his analysis of the generic
surface kinematics of the parametrized classical field theory
\cite{Dirac1951}.
%At this point one might object to eliminating Hi but not H - space
%and time are already treated differently, so might as well ...

The rest of this paper is devoted to developing the above new
strategy for canonical quantum gravity, with emphasis on its
physical interpretation and geometrical basis. In
section~\ref{parasch}, the previously proposed nonlinear
generalization of
a finite-dimensional physical
system  based on a quantum action principle
\cite{Wang2003, WangKessariIrvine2004} is extended to a
field-theoretical description.
This generalization is motivated by the `implicit nature' of the
gravitational Hamiltonian on the CMC foliation. Before proceeding,
it is necessary to lay out the essential prerequisites for
classical canonical gravity. This is given in the first half of
section~\ref{CGRCMC}, with a focus on the elimination of the
momentum constraints leading to a reduced system of the pattern
discussed in section~\ref{parasch}. The second half of
section~\ref{CGRCMC} addresses the quantization of the
`CMC-reduced' canonical general relativity in
section~\ref{CGRCMC}, using the nonlinear quantization method
developed in section~\ref{parasch}. The corresponding quantum
action generates the quantum evolution of the conformal
three-geometry augmented with two elliptic equations that
determine the scale factor and lapse function. In order to
establish the essential framework of the proposed theory with a
view to more complete treatments, some of the discussion given in
this paper remains formal. As in the development of other quantum
gravity ideas, a `toy model' in an even simpler mathematical
setting will help to grasp the salient features of the new
approach. Of course, one should always bear in mind that the
usefulness of this way of extracting guidance is limited
\cite{MacCallumTaub1972, KucharRyan1989}. Despite this caveat, it
is demonstrated in section~\ref{cosmos} that the Class A Bianchi
cosmologies and the Kantowski-Sachs universe can be treated as
nonlinear quantum minisuperspaces of the proposed theory.
%It is
%expected that other homogeneous cosmologies may be treated
%similarly.
The conclusion with discussion on future work is given
in section~\ref{concl}. Below units of $c = \hbar = 16 \pi G = 1$
are adopted.

\section{Nonlinear quantization of a field theory with an\\ `implicit Hamiltonian'}
\label{parasch}

Let $M$ be a three-dimensional space with coordinates
$x^i\;(i=1,2,3)$. With respect to a preferred time $\tau$ consider
the canonical evolution of a field $\phi$ and its conjugate
momentum $\pi$, each with $n$ components, namely ${\phi}^a =
{\phi}^a(x,\tau)$ and $\pi_a = \pi_a(x,\tau)$ {$(a=1,2,\cdots
n)$}. The time-dependent Hamiltonian density $h = h(\phi,
\phi_{,i}, \cdots, \pi, \pi_{,i}, \cdots, \tau)$ is assumed to be
{\em explicitly and locally} defined. Here and henceforth, the
subscript comma and over dot denote differentiation with respect
to a spatial coordinate and time $\tau$ respectively. The
canonical field equations are generated by varying the action
integral
\begin{equation}\label{Sh0}
S[{\phi},\pi] =
\int%_{I}
\!\!\! \int_{M} \left\{ \pi \cdot \dot{\phi} - h \right\}\d^3 x
\,\d\tau
\end{equation}
where $\pi \cdot \dot{\phi} := \pi_a \,\dot{\phi}^a$,
with respect to $\pi$ and $\phi$ to be
\begin{equation}\label{dphi0}
\dot{\phi} = \fd{\HH\;}{\pi}\,,\;
\dot{\pi}  = -\fd{\HH\;}{\phi}
\end{equation}
in terms of the Hamiltonian
\begin{equation}\label{Shh}
\HH = \int_{M}\! h \,\d^3 x .
\end{equation}

%Note that the above Hamiltonian is explicitly defined.
The treatment of a canonical field theory using an implicit
Hamiltonian can be developed as follows. In terms of an arbitrary
time $t$ an action equivalent to \eqref{Sh0} is
\begin{equation}\label{Sh}
S[{\phi},\pi, \sigma, N]  =
\int%_{I}
\!\!\!
\int_{M}
\left\{
{\pi} \cdot \partial_t{\phi}
-
\sigma\, \partial_t{\tau}
-
N \H
\right\}\d^3 x \,\d t
\end{equation}
where
\begin{equation}\label{H=-e+h}
\H = -\sigma + h
\end{equation}
and $\tau = \tau(t)$
together with two Lagrangian multipliers
$\sigma = \sigma(x,t)$ and  $N = N(x,t)$. Note that $\H$
has no explicit dependence on the parameter time $t$. The integral
\begin{equation}\label{}
\Sigma =
\int_{M}
\!
\sigma
\,\d^3 x .
\end{equation}
is introduced for later use.

Now put $t=\tau$ but retain $\tau$ and $N$. The action \eqref{Sh} becomes
\begin{equation}\label{Shhh}
S[{\phi},\pi, \sigma, N] =
\int%_{I}
\!\!\!
\int_{M}
\left\{
\pi \cdot \dot{\phi} - \sigma - N \H
\right\}\d^3 x \,\d\tau .
\end{equation}
This form suggests a generalization from the restricted form of
$\H$ in \eqref{H=-e+h} a generic $\H =$\break
$\H(\phi, \phi_{,i},\cdots, \pi, \pi_{,i}, \cdots, \sigma, \sigma_{,i}, \cdots, \tau)$.
For the purpose of the subsequent discussion, it suffices to
assume that this expression is also explicitly and locally
defined.
With this generic $\H$, the variations of the action \eqref{Shhh}
in $\pi, {\phi}, N, \sigma$ generate, respectively, the equations
\begin{equation}\label{dphi}
\dot{\phi} = \fd{H}{\pi}\,,\;
\dot{\pi}  = -\fd{H}{\phi}
\end{equation}
\begin{equation}\label{H=0}
\H = 0
\end{equation}
\begin{equation}\label{1+H}
1+\fd{H}{\sigma} = 0
\end{equation}
where
\begin{equation}\label{}
H =
\int_{M}
\!
N \H
\,\d^3 x .
\end{equation}
Equations \eqref{dphi} govern the evolution of ${\phi}$ and $\pi$.
Through $H$ these equations are coupled with \eqref{H=0} and
\eqref{1+H} that can be regarded as algebraic or elliptic
equations determining $\sigma$ and  $N$ respectively. For $\sigma$
to be interrelated as the `energy density' of the system, it
should be nonnegative. Clearly,  the above equations reduce to to
\eqref{dphi0} for the `simple case' where \eqref{H=-e+h} holds,
with \eqref{H=0} and  \eqref{1+H} yielding $\sigma = h$ and $ N
=1$.
Given a generic $\H$,
although in principle \eqref{H=0} may be solved for the `true'
nonvanishing Hamiltonian density $\sigma$, this will bring about a high
degree of nonlocality into the reduced canonical theory.
Such an implicit and nonlocal
Hamiltonian may in turn cause
severe conceptual and technical problems for quantization.

The standard canonical quantization of the system with the action
\eqref{Sh0} yields the \sch{} equation
\begin{equation}
\im \dot{\Psi} = \HHHhat \,\Psi \label{scheq}
\end{equation}
for the state functionals $\Psi = \Psi[\phi]$
in terms of the Hamiltonian operator
\begin{equation}\label{Shh}
\HHHhat = \int_{M}\! \hhat \,\d^3 x .
\end{equation}
where $\hhat$ is obtained by substituting $\pi \rightarrow
\hat{\pi} = -\im \fd{}{\phi}$ into $h$ with a suitable choice of
factor ordering. The \sch{} equation \eqref{scheq} can be shown to
follow from the quantum action integral
\cite{WangKessariIrvine2004, DeWitt1964}
\begin{equation}\label{schact}
\SQ[\Psi, \Psi^*]
=
\int\!\left\{\Re \ang{\Psi}{\im\dot\Psi}
-
\ang{\Psi}{\HHHhat \,\Psi} \right\}\,\d \tau
\end{equation}
using the inner product
of any state functionals $\Psi_1$ and $\Psi_2$
in terms of the functional integration
\begin{equation}
\label{angPsi} \ang{\Psi_1}{\Psi_2} := \int \Psi_1^* \Psi_2
\,\D \phi
\end{equation}
in an appropriate sense. The operator $\hhat$ and other physical
observables are assumed to be Hermitian with respect to this inner
product. The expectation value of any Hermitian operator $\hat{O}$
with respect to $\Psi$ is defined, in a
standard manner, as:
\begin{equation}\label{aveO}
\av{\hat{O}} := \frac{\ang{\Psi}{\hat{O}\,\Psi}}{\ang{\Psi}{\Psi}} .
\end{equation}

By the same token as the steps leading from \eqref{Sh0} to \eqref{Sh},
an action equivalent to \eqref{schact} can be written, using the
Lagrangian multipliers
$\sigma(x,t)$ and  $N(x,t)$,
as
\begin{equation}\label{schact1}
\SQ[\Psi, \Psi^*, \sigma, N]
= \int\!\left\{ \Re \ang{\Psi}{\im \dot\Psi}
-
\Sigma \ang{\Psi}{\Psi}
-
\ang{\Psi}{\HHhat \,\Psi}
\right\}\d\tau
\end{equation}
where
\begin{equation}\label{HHhat}
\HHhat =
\int_{M}
\!
N \Hhat
\,{\d^3 x}
\end{equation}
with
\begin{equation}\label{QH=-e+h}
\Hhat = -\sigma + \hhat .
\end{equation}

Here $\Hhat$ is the operator of $\H$ given in \eqref{H=-e+h}. For
a generic $\H =$ $\H(\phi, \phi_{,i}, \cdots, \pi, \pi_{,i},
\cdots, \pi, \sigma, \sigma_{,i}, \cdots)$ as discussed above that
admits a corresponding Hermitian operator $\Hhat$, the use of this
operator in the quantum action \eqref{schact1} will, under
variations with respect to $\Psi$, its conjugate,  $N$ and
$\sigma$, generate the following generalized form of quantum
evolution equations
\begin{equation}\label{paramscheq}
\im\dot \Psi = \HHhat \Psi
%\im\dot \Psi = \HHhat \Psi
\end{equation}
\begin{equation}\label{paramH}
\av{\Hhat} = 0
\end{equation}
\begin{equation}\label{param1+H}
1+\ave{\fd{\HHhat}{\sigma}} = 0
\end{equation}
respectively,
up to a (physically irrelevant) overall time dependent phase of
$\Psi$.

The \sch{} type equation \eqref{paramscheq} thus describes
the {\em unitary} quantum evolution of $\Psi[\phi]$. The
`Hamiltonian operator' $\HHhat$ depends on $\sigma$ and  $N$, for
which \eqref{paramH} and \eqref{param1+H}, must be solve. As these
two equations are nonlinear in the state functional, the overall
quantum evolution is {\em nonlinear}. Although the introduction of
nonlinearity into quantum gravity is relatively new, the nonlinear
modification of quantum mechanics has long been receiving
considerable attention in addressing the measurement problem
\cite{Pearle1976, Laloe2001, BassiGhirardi2003}. Interestingly,
the energy incurred by the nonlinear term in certain `dynamical
reduction models' can be shown to be comparable to the
gravitational self-energy of the measurement apparatus
\cite{Pearle1976}. This leads one to suspect that gravity might be
related to quantum evolution and measurement in a nonlinear
manner. For further discussion on nonlinear quantum theories, see
e.g. \cite{Mielnik1974, KibbleRandjbarDaemi1980, Weinberg1989}. In
view of this, the nonlinear quantum formalism presented in this
section has been developed with a view to quantizing general
relativity.  It will be demonstrated in the next section that the
action integral for classical canonical gravity does indeed reduce
to the form of \eqref{Shhh} under the CMC foliation. The nonlinear
quantization scheme may therefore be employed for quantum gravity
as discussed below.

\section{Canonical general relativity on the CMC foliation and its nonlinear quantization}
\label{CGRCMC}

In this section the {essentials} of the Dirac-ADM formulation of
canonical general relativity {are} first recapitulated. The theory
is then gauge-fixed by choosing the spatial hypersurfaces to
possess a constant mean exterior curvature. The resulting
`CMC-reduced' formulation is readily organized into the
form of the generalized canonical field theory discussed in the
forgoing section that may be quantized in a nonlinear fashion.

Under the ADM 3+1 split of spacetime, each spatial hypersurface
arbitrarily  labelled by time $t$ has a spatial metric
$g$ with components $g_{i j}$ in coordinates $x^i$ $(i=1,2,3)$.
The inverse (i.e. contravariant) metric $g^{-1}$ has components
$g^{i j}$.
In terms of the lapse function $N$ and shift vector $X$
with components $X^i$, the spacetime line element takes the form
\begin{equation}\label{metric}
\d s^2 = -N^2\, \d t^2 + g_{i j} \,(\d x^i+X^i\,\d t) (\d x^j+X^j\,\d t) .
\end{equation}
Here indices are lowered or raised using $g_{i j}$ and $g^{i j}$
respectively.

The extrinsic curvature tensor $K$ has components given by
\begin{equation}\label{Kij}
K_{i j} = \frac1{2N}\left( -\dot{g}_{i j} +  X_{i;j} +  X_{j;i}  \right)
\end{equation}
where the semicolon denotes a covariant derivative using the
Levi-Civita connection of $g$. This connection gives rise to the
intrinsic scalar curvature $\R$. Of particular importance for the
present discussion is the mean exterior curvature defined as the
trace of $K$  by
\begin{equation}\label{MeanK}
\K := g^{ij} K_{ij} %= \frac12\, \mu^{-1} g_{ij}\, p^{ij} .
\end{equation}

In terms of $K$ and $\K$ the `metric momentum' tensor density
$p$ is given by
\begin{equation}\label{pij}
  p^{i j} = -  \mu\left(K^{i j} - g^{i j} \K \right)
\end{equation}
where  $\mu = (\det g)^\frac12$ is the  scale factor for the
spatial volume element.

The well-known Dirac-ADM action integral for
canonical gravity takes the form
\begin{equation}\label{S}
S[g, p, N, X] =
\int%_{I}
\!\!\!
\int_{M}
\left\{
p \cdot \dot{g}
- N \H
- X \cdot \J
\right\}\d^3 x \,\d t
\end{equation}
where $p \cdot \dot{g} = p^{ij} \dot{g}_{ij}$, $X \cdot \J = X^i
\J_i$ in terms of the Hamiltonian and momentum constraints $\H$
and $\J_i$ given by
\begin{equation}
\label{Hconstr}
\H
=
%G_{i j k l}
\frac1{2}\,\mu^{-1} (g_{i k}g_{j l} + g_{i l}g_{j k} -  g_{i
j}g_{k l} )\, p^{i j} p^{k l} -  \R \mu
\end{equation}
\begin{equation}
\label{Jconstr}
\J_i
=
-2 g_{ij} \, p^{j k}{}_{;k}
%-2 g_{ij} \nabla_k p^{j k}
=
-2\,g_{ij}\,p^{jk}{}_{,k}
-2\, g_{ij,k}\,p^{jk}
+ g_{jk,i}\,p^{jk}
\end{equation}
respectively.

The roles of $N$ and $X^i$ as Lagrangian
multipliers result in the Hamiltonian constraint equation
\begin{equation}\label{H}
\H = 0
\end{equation}
and the momentum constraint equations
\begin{equation}\label{Ha}
\J_i = 0 .
\end{equation}

York's treatment of canonical gravity begins by
expressing the metric $g$ as
\begin{equation}\label{gphi}
g = \varphi^{4} \gamma
\end{equation}
in terms of a conformal factor $\varphi$ and
a conformal metric $\gamma$
satisfying a `normalization' condition.
This may be done by
fixing its volume scale factor $\mu_\gamma = (\det \gamma)^\frac12$
\cite{ChoquetBruhatYork1980, Kuchar1992a}
or its scalar curvature $\R_{\gamma}$ whereby regarding $\gamma$ as a
Yamabe metric \cite{Yamabe1960, Schoen1984, FischerMoncrief1997}. E.g.
\begin{equation}\label{mu_gamma1}
\mu_\gamma = 1
\end{equation}
or
\begin{equation}\label{R_gamma_fix}
\R_\gamma = \pm1, 0
\end{equation}
depending on the topology of the spatial hypersurface $M$. The
present discussion does not depend explicitly on the choice of
such normalization conditions. However, boundary terms in various
integrations over $M$ are dropped at will for the sake of
mathematical convenience. That is to say, either a compact $M$ or
an adequate fall-off of the relevant fields over $M$ at infinity
is assumed.

An interesting result of York's is that the momentum constraints
decouples from the Hamiltonian constraint under the CMC foliation.
Furthermore, the momentum constraint equations  \eqref{Ha} are
satisfied if and only if the longitudinal part of the metric
momentum $p$ vanishes, leading to the expression
\begin{equation}\label{pphi}
p = \varphi^{-4}\varpi + \frac23\,\varphi^{2}\K\, \gamma^{-1} \mu_\gamma
\end{equation}
where $\varpi$ is a transverse traceless tensor density with
respect to $\gamma$ \cite{York1971, York1973}.

The Hamiltonian constraint equation gives rise to the Lichnerowicz
equation \cite{Lichnerowicz1944, York1972, FischerMoncrief1997,
Kuchar1992a}
\begin{equation}\label{Leq}
 \Delta_{\gamma} \varphi
-\frac3{64}\, \tau^2 \varphi^5
-\frac18\,\R_{\gamma} \varphi
+\frac18\,{\gamma_{ik} \gamma_{jl}}\,\varpi^{ij}\varpi^{kl}\, \mu_\gamma{}^{\!\!\!-2} \,\varphi^{-7}
= 0
\end{equation}
regarded as the elliptic equation for the conformal factor $\varphi$
if \eqref{H}
is also to be solved. The present analysis will however defer this process
by retaining $\H$ in the action. Since
\begin{equation}\label{}
\mu = \varphi^{6} \mu_\gamma
\end{equation}
in the subsequent discussion the conformal factor $\varphi$ will
be eliminated using
\begin{equation}\label{}
\varphi = \mu_\gamma{}^{\!\!-\frac16}\mu^{\frac16} .
\end{equation}
Following \cite{FischerMoncrief1997, FischerMoncrief1999} one
writes
\begin{equation}\label{}
p \cdot \dot{g} =  \varpi \cdot \dot{\gamma} + \frac43\,\K\,  \dot \mu
\end{equation}
%$\dot{\;} = \partial_t$
where $\varpi \cdot \dot{\gamma} = \varpi^{ij}\, \dot {\gamma}_{ij}$. This,
together with the fact that \eqref{Ha} has been solved by \eqref{pphi}
and an integration by parts,
reduces the Dirac-ADM action \eqref{S} to the form
\begin{equation}\label{SS}
S[\gamma, \varpi, \mu, \K, N] =
\int%_{I}
\!\!\!
\int_{M}
\left\{
 \varpi \cdot \dot{\gamma}
-
\frac43\,\mu\,\dot \K
- N \H
\right\}\d^3 x \,\d t .
\end{equation}
Here the CMC-reduced Hamiltonian constraint
%$\H = \H(\gamma, \varpi, \mu, \kappa)$ given by is
{$\H = \H(\gamma, \varpi, \mu, \tau)$ is}
obtained by substituting \eqref{gphi} and \eqref{pphi} into
\eqref{Hconstr} and has the form
\begin{eqnarray}\label{HR}
\H &=& -\frac38\, \mu \tau^2 +\mu^{-1} {\gamma_{ik}
\gamma_{jl}}\,\varpi^{ij}\varpi^{kl} -  \R \mu\\
\label{HCMC}
&=&
-\frac38\, \mu \tau^2
+
\mu^{-1} {\gamma_{ik}\gamma_{jl}}\,\varpi^{ij}\varpi^{kl}
- \R_{\gamma}\mu_\gamma{}^{\!\frac23} \mu^{\frac13}
+ 8\,\mu_\gamma{}^{\frac56}\mu^{\frac16} \,
\Delta_{\gamma}(\mu_\gamma{}^{\!\!-\frac16}\mu^{\frac16})
\end{eqnarray}
where
$\Delta_{\gamma}$
is the Laplace-Beltrami operator on scalar functions with respect to
$\gamma$
%with the associated Levi-Civita connection $\nabla_{\!\gamma}$
(following the sign convention of \cite{ChoquetBruhatYork1980, Kuchar1992a}
as opposed to that of
\cite{York1973, FischerMoncrief1997}). Note that the action \eqref{SS} has a
structure analogous to that of
\eqref{Sh}. To bring \eqref{SS} to the form of
\eqref{Shhh}, where $\sigma$, $\phi$ and $\pi$ may be
identified as $\mu$, $\gamma$ and $\varpi$ respectively,
the time parametrization $t = \tau := \frac43\K$ is adopted, yielding
the action
\begin{eqnarray}\label{SS1}
S[\gamma, \varpi, \mu, N]
&=&
\int%_{I}
\!\!\!
\int_{M}
\left\{
\varpi \cdot \dot{\gamma}
-
\mu
-
N \H
\right\}
\d^3 x \,\d\tau
\\
\label{SS2}
&=&
\int%_{I}
\!
\left\{
\int_{M}
\!\!
\varpi \cdot \dot{\gamma}
\,\d^3 x
-
\vol
-
H
\right\}
\d\tau
\end{eqnarray}
where
\begin{equation}\label{h}
H =\int_{M} N \H \,\d^3 x
\end{equation}
and
\begin{equation}\label{}
\vol = \int_M \mu \,\d^3 x
\end{equation}
representing the `volume of the universe'. It follows that

\begin{eqnarray}
\label{fdHmu}
\fd{H}{\mu}
&=&
\frac43\,\Delta N
-
N\!\left\{\frac38\,\tau^2 +\mu^{-2} {\gamma_{ik}
\gamma_{jl}}\,\varpi^{ij}\varpi^{kl} + \frac13 \, \R
\right\}
\\
&=&
\frac43\, \mu_\gamma{}^{\frac56}\mu^{-\frac56} \,\Delta_{\gamma} (N \mu_\gamma{}^{\!\!-\frac16}\mu^{\frac16})
\nonumber
\\
\label{dkappa1}
&&
\hspace{-8pt}
-
N\!\left\{
\frac38\,
%\kappa^2
{\tau^2}
+\mu^{-2} {\gamma_{ik} \gamma_{jl}}\,\varpi^{ij}\varpi^{kl}
+\frac13\,  \R_{\gamma}\mu_\gamma{}^{\!\frac23} \mu^{-\frac23}
-\frac43\,\mu_\gamma{}^{\frac56} \mu^{-\frac56} \,\Delta_{\gamma} (\mu_\gamma{}^{\!\!-\frac16}\mu^{\frac16})
\right\}
\end{eqnarray}
where $\Delta$ is the Laplace-Beltrami operator on scalar
functions with respect to $g$.

In complete analogy with
\eqref{dphi},
\eqref{H=0} and
\eqref{1+H}, the variations of the
the CMC-reduced action \eqref{SS1}
with respect to
$\varpi, {\gamma}, N$ and $\mu$ generate the
canonical type field equations
\begin{equation}\label{dgamma}
\dot{\gamma} = \fd{H}{\varpi}
\,,\;
\dot{\varpi} = -\fd{H}{\gamma}
\end{equation}
for the conformal metric $\gamma$ and its
transverse traceless momentum density $\varpi$,
supplemented with
\begin{equation}\label{HCMC=0}
\H = 0
\end{equation}
and
\begin{equation}\label{1+fdH}
1 + \fd{H}{\mu} = 0
\end{equation}
as the elliptic equations for the scale factor $\mu$ and lapse
function $N$, respectively.

Proceeding with the nonlinear quantization scheme formulated in
section~\ref{parasch}, one now considers a state functional
$\Psi = \Psi_\tau[\gamma]$ depending on the conformal metric $\gamma$
modulo diffeomorphisms \cite{Carlip2001}. The inner product of any
two such state functionals ${\Psi_1}$ and ${\Psi_2}$ are given in
terms of an appropriate functional integration of the form
\begin{equation}
\label{angPsi}
\ang{\Psi_1}{\Psi_2}
:=
\int \Psi_1^* \Psi_2
\,\D \gamma
\end{equation}
over the equivalent classes of
the conformal metric $\gamma$ related also {by
%modulo
diffeomorphisms.} Corresponding to the classical action
\eqref{SS2}, one further writes down the quantum action for the
above $\Psi_\tau[\gamma]$, $\mu(x,\tau)$ and $N(x,\tau)$ as
\begin{equation}\label{scheq2}
\SQ[\Psi, \Psi^*, \mu, N]
= \int\!\left\{ \Re \ang{\Psi}{\im \dot\Psi}
-
\vol \ang{\Psi}{\Psi}
-
\ang{\Psi}{\HHhat \,\Psi}
\right\}\d \tau .
\end{equation}
This action in turn generates the following \sch{} type equation
\begin{equation}\label{Qdgamma}
\im\dot \Psi = \HHhat \Psi
\end{equation}
for $\Psi$ (up to an overall time dependent phase),
coupled with the elliptic type equations
\begin{equation}\label{QHCMC=0}
\av{\Hhat} = 0
\end{equation}
\begin{equation}\label{Q1+fdH}
1+\ave{\fd{\HHhat}{\mu}} = 0
\end{equation}
which serve to determine $\mu$ and $N$, respectively.
In the above equations, the operator $\Hhat$ is obtained by
substituting
$\varpi \rightarrow \hat\varpi := -\im \fd{}{\varpi}$
into
\eqref{HCMC} where the term quadratic in $\varpi$
naturally becomes the Laplace-Beltrami operator
in the space of conformal three-geometries.
It is worth noting that the above procedure
still carries through if the conformal factor $\varphi$
is chosen in place of $\mu$, with \eqref{Q1+fdH} slightly
modified as $\mu$ becomes an operator in terms of $\gamma$.

The \sch{} type equation \eqref{Qdgamma} provides a unitary
quantum description of the conformal three-geometry. Through the
transverse and traceless nature of the corresponding momenta, it
is gravitational waves that is quantized. The quantum evolution
depends on $\mu$ and $N$ which are determined by the elliptic type
equations \eqref{QHCMC=0} and \eqref{Q1+fdH}. These two equations
make the overall quantum evolution nonlinear and nonlocal in the
the state functional. Nevertheless, the Hamiltonian constraint
operator is locally defined. Its kinetic part is quadratic in the
momentum operator and should be represented by the Laplacian in
the space of conformal three-geometries.

\section{Nonlinear quantum minisuperspaces}
\label{cosmos}

The nonlinear framework for quantum gravity presented in the
previous section is complicated by the presence of the Laplacian
of $\mu$ and $N$ as well as the field-theoretical nature of the
problem. This complexity can be avoided in the cosmological
approach with finite degrees of freedom. At this oversimplified
but concrete level, one expects to gain intuition about the full
theory for being able to carry out explicit analysis.

For this purpose, the Class A Bianchi type models and
Kantowski-Sachs universe \cite{RyanShepley1975,
StephaniKramerMacCallumHerlt2003} will be treated as nonlinear
quantum minisuperspaces of the full theory under discussion. These
cosmological models are described by the lapse function $N(t)$,
length scale factor $R(t)$ as well as two additional functions
$\beta_{+}(t)$ and $\beta_{-}(t)$  to allow for the dynamical
anisotropy of the spacial hypersurface at any parameter time $t$.
The spacetime line element takes the following common form
\begin{equation}\label{ds2}
\d s^2=-N^2 \d t^2 + R^2(e^{2 \beta})_{ij} \,e^i
 e^j
\end{equation}
for $i,j = 1,2,3$, where $e^i$ denote some basis 1-forms of the
spatial hypersurface and $\beta$ is a traceless matrix  with
elements given by
\begin{equation}
\beta_{ij}=\diag(
\beta_{+} + \sqrt{3}\beta_{-},\,
\beta_{+} - \sqrt{3}\beta_{-},\,
-2\beta_{+}
) .
\end{equation}
The basis 1-forms $e^i$ satisfy the structure equation
\begin{equation}\label{}
\d e^i = \frac{1}{2}\,C^i{}\!_{jk}~e^j \wedge e^k
\end{equation}
with certain model-specific  structure constants
$C^i{}\!_{jk}$ {\cite{RyanShepley1975}}.
Evidently, the spatial metric $g$ may be written as
\begin{equation}\label{}
g = R^2 \gamma
\end{equation}
where
\begin{equation}\label{}
\gamma_{ij} = (e^{2 \beta})_{ij}
\end{equation}
satisfying the normalization condition \eqref{mu_gamma1}.
Hence $\mu = R^3$.
% scale factor $\mu = R^3$
%$p^{ij}$ has no longitudinal part due to homogeneity.

It follows from \eqref{Kij}, \eqref{MeanK}, \eqref{pij} and
\eqref{pphi} that
\begin{equation}\label{}
\tau = \frac43\K = -\frac{4 \dot R}{N R}
\end{equation}
{with the over dot indicating differentiation with respect to
$t$,} and
\begin{equation}\label{}
\varpi^{ij}
=
\frac{R^3}{N}\,
\diag\left[
(\dot\beta_{+} + \sqrt{3}\dot\beta_{-})\,e^{-2\beta_{+} -2 \sqrt{3}\beta_{-}},\,
(\dot\beta_{+} - \sqrt{3}\dot\beta_{-})\,e^{-2\beta_{+} +2 \sqrt{3}\beta_{-}},\,
-2\dot\beta_{+}\,e^{4\beta_{+}}
\right] .
\end{equation}
Therefore
\begin{equation}\label{}
\varpi \cdot \dot {\gamma}
=
\frac{12\mu}N\,(\dot\beta_{+}^2+\dot\beta_{-}^2)
=
\varpi_+\dot\beta_{+}+\varpi_-\dot\beta_{-}
%=:
%\varpi_\pm\dot\beta_\pm
\end{equation}
where
\begin{equation}\label{}
\varpi_{\pm} = \frac{12\mu}{N}{\dot{\beta}_{\pm} .}
\end{equation}

With the choice of time $t = \tau$ and the substitution (as simple
volume normalization) $\int_M\!\d^3 x  \rightarrow 1$ the action
\eqref{SS1} reduces to
\begin{equation}\label{BSS}
S[{\beta}_{\pm}, \varpi_{\pm}, \mu, N] =
\int\!
\left\{
\varpi_+\dot\beta_{+}+\varpi_-\dot\beta_{-}
-\mu - N \H
\right\}\d\tau
\end{equation}
where
\begin{equation}
\label{HHB}
\H
=
%- \frac{\Pi^2}{24 R}
-\frac38\, \mu \tau^2
%+\frac{1}{24 R^3}\,({p^2_+ +p^2_-})
%+\frac{1}{24 \mu}\,({p^2_+ +p^2_-})
+\frac{\varpi^2}{24 \mu}
%+R\,V_\beta
-\mu^{\frac13}\,\R_\gamma
\end{equation}
with $\varpi^2 := {\varpi^2_+ +\varpi^2_-}$.

As no factor ordering ambiguity arises,
the cosmological model may now be
quantized using the
standard substitutions: $\varpi_\pm \rightarrow \hat{\varpi}_\pm = -\im \partial_{\beta_\pm}$
and
%${p}^2_+ + {p}^2_- \rightarrow -\partial_+^2-\partial_-^2 = -\Delta_\beta$
%$\hat{p} \rightarrow \hat{p}^2 =  -\Delta_\beta = -(\partial_{\beta_+}^2+\partial_{\beta_-}^2)$
${\varpi}^2 \rightarrow \hat{\varpi}^2 = -(\partial_{\beta_+}^2+\partial_{\beta_-}^2)$.
%which is the
%Laplace-Beltrami operator in the 2-dimensional Euclidean space with Cartesian coordinates
%$\beta_\pm$.
These operator are to act on a wavefunction $\Psi = \Psi(\beta_\pm,\tau)$.
The inner product of two such  wavefunctions $\Psi_1, \Psi_2$ {is} simply
\begin{equation}
\label{BangPsi}
\ang{\Psi_1}{\Psi_2} := \int_{-\infty}^{\infty}  \!\int_{-\infty}^{\infty}  \!
\Psi_1^*\, \Psi_2 \,\d\beta_+\d\beta_- .
\end{equation}

As per preceding discussions,  the quantum action of \eqref{BSS} can be written as
\begin{equation}\label{scheq3}
\SQ[\Psi, \Psi^*,\mu, N]
= \int\!\left\{
\Re \ang{\Psi}{\im \dot\Psi}
-
\mu\ang{\Psi}{\Psi}
-N  \ang{\Psi}{  \Hhat \,\Psi} \right\}\d \tau .
\end{equation}
This generates the \sch{} type equation
\begin{equation}\label{Bsch}
{\im \dot{\Psi}}
=
N\!\left\{
\frac{\hat{\varpi}^2}{24 \mu}
-\mu^{\frac13}\,\R_\gamma
\right\}  {\Psi}
\end{equation}
for $\Psi(\beta_\pm,\tau)$ (up to an overall time dependent phase), and
following two equations
\begin{equation}\label{BH0}
-\frac38\, \mu \tau^2 + \frac{\av{\hat{\varpi}^2 }}{24 \mu} -\mu^{\frac13} \,\av{\R_\gamma} = 0
\end{equation}
\begin{equation}\label{BH1}
1-N\left\{
\frac38\, \tau^2 + \frac{\av{\hat{\varpi}^2 }}{24 \mu^2} + {\frac13}\, \mu^{-\frac23} \,\av{\R_\gamma}
\right\} = 0
\end{equation}
algebraic in $\mu(\tau)$ and $N(\tau)$. Equation \eqref{Bsch}
describes the unitary quantum evolution of the anisotropy (i.e.
`truncated gravitational waves') of the cosmological model in a
mechanical fashion. In particular, it involves a `Hamiltonian
operator' consisting of the sum of a kinetic term and a potential
term. However, the presence of the supplementary conditions
\eqref{BH0} and \eqref{BH1} introduces nonlinearity as well as
nonlocality into the description. In this manner, equations
\eqref{Bsch}--\eqref{BH1} constitute a consistent system of {\em
nonlinear integro-partial differential equations}. Such equations
can be solved at least numerically in a fashion similar to
\cite{Wang2003} where a computational approach to the nonlinearly
quantized Friedmann universe is detailed. In the case of the
Bianchi I model, the properties $\R_\gamma=0$ and
$\av{\hat{\varpi}^2 }$ being a constant enable one to find exact
solutions to \eqref{Bsch}--\eqref{BH1} using the analytical method
developed in \cite{WangKessariIrvine2004}.

%\section{Conclusion and outlook}
\section{Concluding remarks}
\label{concl}

A discussion has been given of a nonlinear quantization scheme for
canonical field theories with an implicitly defined Hamiltonian.
Two Lagrangian multipliers are {involved} in the formulation: one
representing the value of the energy density and the other
effecting this value as an implicit function of the canonical
variables allowing for explicit time-dependence. This structure
has been shown to naturally arise from canonical general
relativity on the constant mean curvature foliation, providing a
basis for a nonlinear quantum theory of gravity. As the scale
factor (or conformal factor) and lapse function remain
unquantized, they act as Lagrangian multipliers in a quantum
action principle. In particular, the scale factor represents the
effective positive energy density. The true dynamical degrees of
freedom for general relativity, as identified by York, are carried
by the conformal three-geometries with transverse traceless
momenta and are quantized nonlinearly in the proposed theory.
York's exterior time, namely the constant mean exterior curvature,
or physically the (local) rate of expansion of the universe, is
also adopted here as the preferred time. This work has set the
stage for more explicit and detailed investigations. Indeed, work
is underway to transcribe the present geometrodynamical
description to a `connectodynamical' description so as to
assimilate powerful functional techniques offered by the loop
quantum gravity approach. As the presented framework features
inherent nonlinearity as well as nonlocality in the state
functional, it is of interest to explore the resulting physical
consequences. This will form a subject for future research,
together with other issues such as the inclusion of matter
interaction. On this last point, it is reasonable to anticipate
that much of the essential methodology of the present work for
pure gravity to be carried over, owing to the versatility of the
constant mean curvature analysis in accommodating matter fields.

\section*{Acknowledgments}
I take this opportunity to express gratitude to many colleagues
for enlightening conversations and insightful comments on this
work, especially to Professors Arthur E Fischer, Chris J Isham,
Karel V Kucha\v{r}, Niall O'Murchadha, Robin W Tucker and Dr
Edward Anderson. The research programme has benefitted from partial
support from the EPSRC and Centre for Fundamental Physics, CCLRC.

\end{document}